\documentclass[%
 aip,
 amsmath,amssymb,
 reprint,%
]{revtex4-1}
\usepackage{upgreek}
\usepackage[caption=false]{subfig}
\usepackage{graphicx}
\usepackage{dcolumn}
\usepackage{bm}

\usepackage[utf8]{inputenc}
\usepackage[T1]{fontenc}
\usepackage{mathptmx}
\usepackage{etoolbox}

\makeatletter
\def\@email#1#2{%
 \endgroup
 \patchcmd{\titleblock@produce}
  {\frontmatter@RRAPformat}
  {\frontmatter@RRAPformat{\produce@RRAP{*#1\href{mailto:#2}{#2}}}\frontmatter@RRAPformat}
  {}{}
}%
\makeatother
\begin{document}

\preprint{AIP/123-QED}

\title[]{Ultra-high precision comb-locked terahertz frequency-domain spectroscopy of whispering-gallery modes}
\author{Sebastian M\"uller}
 \altaffiliation[Also at ]{Institute of Microwaves and Photonics, Friedrich Alexander University, Erlangen, Germany}
 \affiliation{TOPTICA Photonics AG, Lochhamer Schlag 19, 82166 Munich, Germany}

\author{Kane Hill}
 \altaffiliation[Also at ]{Te Whai Ao Dodd-Walls Centre for Photonic and Quantum Technologies, New Zealand}
 \affiliation{Department of Physics, The University of Auckland, Auckland 1010, New Zealand}

\author{Dominik Walter Vogt}
 \altaffiliation[Also at ]{Te Whai Ao Dodd-Walls Centre for Photonic and Quantum Technologies, New Zealand}
 \affiliation{Department of Physics, The University of Auckland, Auckland 1010, New Zealand}

\author{Thomas A. Puppe}
 \affiliation{TOPTICA Photonics AG, Lochhamer Schlag 19, 82166 Munich, Germany}
    \email{thomas.puppe@toptica.com}
 \author{Yuriy Mayzlin}
 \affiliation{TOPTICA Photonics AG, Lochhamer Schlag 19, 82166 Munich, Germany}

 \author{Rafal Wilk}
 \affiliation{TOPTICA Photonics AG, Lochhamer Schlag 19, 82166 Munich, Germany}
 

\date{\today}

\begin{abstract}
We demonstrate the capabilities of a novel frequency-domain terahertz spectrometer based on a comb-locked frequency synthesizer, which provides absolute frequency calibration. The inherent stability and repeatability of the scans allow for the combination of fast data acquisition with an average time-limited signal-to-noise ratio. We demonstrate kilohertz level frequency resolution in terahertz precision spectroscopy of ultrahigh quality whispering-gallery-mode resonators. Spectra covering multiple free spectral ranges (>36GHz) with sub-20kHz resolution are acquired in 5s. We analyse the coupling behaviour and temperature tuning of single resonances and, for the first time, observe minute red and blue shifts of different mode families. The experimental results are supported with finite element simulations. 
\end{abstract}

\maketitle

\section{Introduction}

Frequency-domain spectrometers (FDS) based on optoelectronic terahertz (THz) generation driven by the difference frequency of two optical fields transfer the tunability and precision of the lasers to the THz domain \cite{nellen2018recent}. State-of-the-art FDS rely on the frequency stability and phase-noise of two independent laser sources due to the difficulty of establishing a lock between sources separated by multiple GHz. Moreover, several tunable distributed feedback (DFB) lasers are combined to take full advantage of the bandwidth of current photo mixers \cite{Deninger:15, lu2022ultrafast}. Preserving a lock during scanning over hundreds of GHz is even more challenging. The system presented here is based on an external endless frequency-shifter of the optical comb spectrum of a short-pulse erbium-fiber oscillator \cite{Telle:10,Rohde:14}. It establishes phase-predictable tuning of a continuous-wave (CW) external-cavity diode laser at speeds up to 1 THz/s and over spectral bands as large as 10 THz suitable for precision broadband IR spectroscopy \cite{Gotti_2020}. This optical synthesizer is combined with a second fixed-frequency phase-locked laser, thus enabling comb-referenced generation and coherent detection of CW THz radiation by (difference) frequency mixing \cite{RobinsonTait:2019}. This system has the potential to enable precision measurement equipment to support the development of 6G terahertz photonics \cite{Kang2024, eichler2023}.  

\section{Spectrometer and experimental setup}

\begin{figure}[t!]
\centering\includegraphics[width=0.45\textwidth]{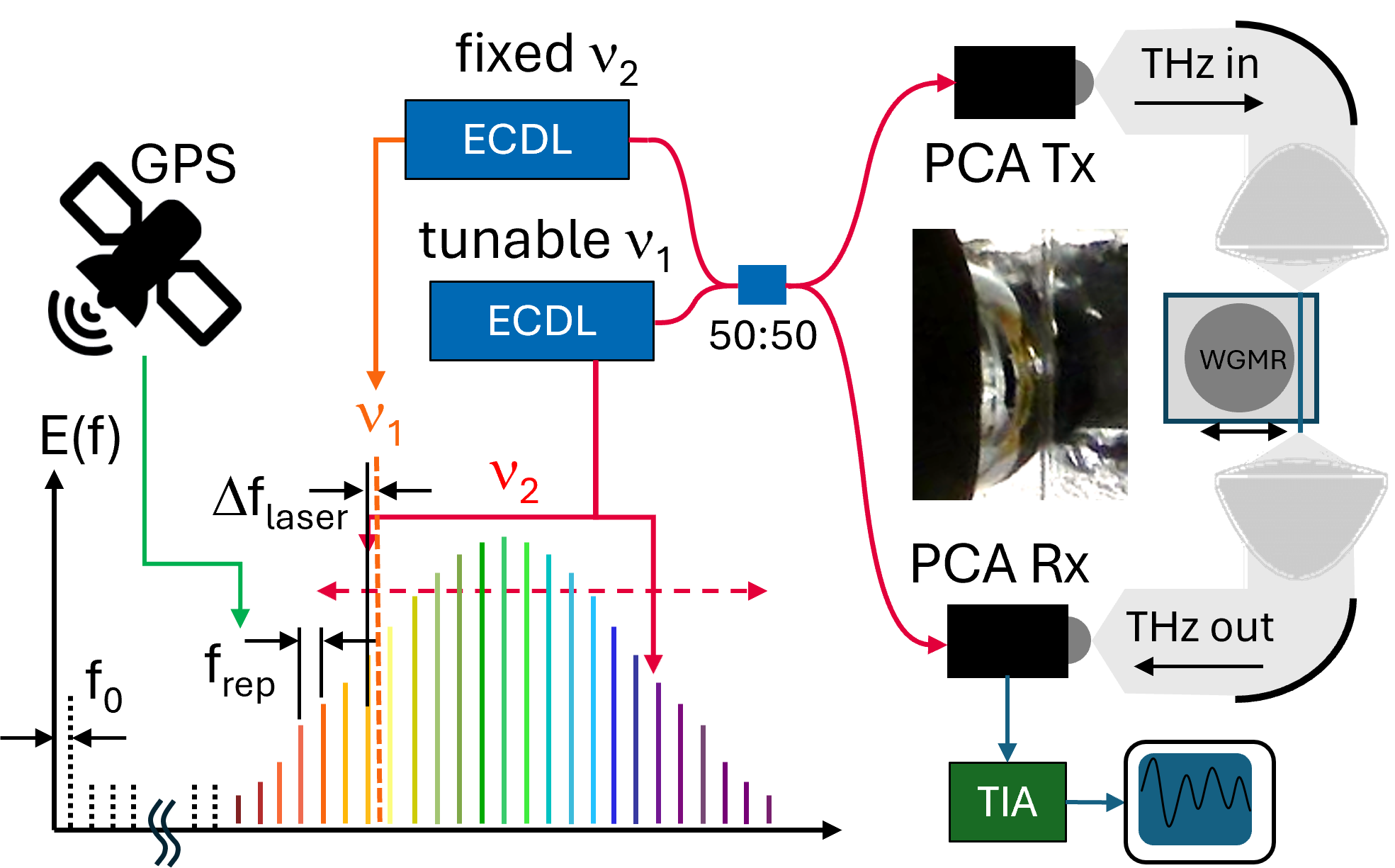}
\caption{Simplified schematic showing the principle of the novel THz FDS as well as the experimental setup and the data acquisition using a trans-impedance amplifier (TIA) and a high-resolution (12bit)/high-speed sampling oscilloscope. The photograph shows the mounted 4\,mm diameter silicon sphere next to the air-silica step-index waveguide. Two external-cavity diode lasers (ECDL) are locked to a common f\textsubscript{rep} stabilised frequency comb. The external frequency shifter allows for tuning of one of the phase-locked ECDLs ($\nu_1$). The CW light fields are amplified and combined in a 50/50 fibre splitter to drive the frequency-domain THz spectrometer shown on the right.}
\label{fig:1}
\end{figure}

The laser engine of the THz spectrometer is shown in \mbox{Fig. \ref{fig:1}}. Two CW external-cavity diode lasers (ECDL) are phased-locked to a common frequency comb spectrum generated by a mode-locked erbium-fiber oscillator. Within the locking bandwidth ($\sim$1MHz), they inherit the phase-noise properties of the comb lines. Since the THz signal is based on difference-frequency generation, it is sufficient to lock the repetition rate f\textsubscript{rep}=200 MHz to a low-noise radio-frequency (RF) reference oscillator. By locking the RF oscillator to a GPS-disciplined oscillator at 10 MHz, the repetition rate f\textsubscript{rep} is referenced to the SI second, allowing for reproducible measurements across different laboratories. Both CW lasers ($\nu_1, \nu_2$) are phase-locked to its neighbouring comb mode at a fixed RF offset frequency derived from the repetition rate f\textsubscript{rep}. To enable broadband tuning of one of the lasers ($\nu_1$) while preserving the phase lock and realising an optical-frequency synthesizer (OFS), the comb spectrum is shifted by an external endless frequency-shifter \cite{Telle:10}. The frequency shifter is based on serrodyne shifting of the instantaneous carrier frequency by applying a corresponding time-dependent phase shift with an electro-optic modulator (EOM) synchronised to the pulse train. It takes advantage of the $2\pi$ periodicity of the optical phase, applying the $2\pi$ fly-backs in between pulses, greatly reducing spurious signals. Details on external endless frequency-shifter can be found in \cite{Rohde:14,Puppe:xx}. The ECDL ($\nu_1$) is continuously tunable without any mode-hops over 100\,nm. Adapting the spectral width of the comb allows tuning over the common spectral range. The tuning speed of up to $>1$~THz/s has been demonstrated for a predetermined frequency sweep limited by the mode-hop-free tuning speed of the CW laser.


We experimentally verify the frequency resolution and absolute frequency stability of the novel FDS by analysing ultra-high quality THz whispering-gallery modes (WGMs). A schematic of the experimental setup is shown on the right-hand side in Fig. \ref{fig:1}. The superimposed light from the FDS based on the OFS is focused onto two photoconductive antennas (PCAs) to generate and detect the coherent THz radiation. Off-axis parabolic mirrors and specifically designed ultra-high weight molecular polyethylene symmetric-pass lenses are used to collimate and focus the THz radiation into a sub-wavelength air-silica step-index waveguide \cite{lo2008aspheric}. With a diameter of 200$\upmu$m, the dielectric waveguide supports single-mode guidance in the frequency range from about 350\,GHz to 700\,GHz. More importantly, with an effective refractive index ${n}_{\mathrm{eff}}$ of about 1.3 at 460\,GHz, the coupling waveguide is phase matched to higher-order radial modes of the investigated whispering-gallery-mode resonator (WGMR). The WGMR is a 4\,mm diameter sphere of high-resistivity silicon with a resistivity $>$10k$\Omega$cm. These resonators are known to support THz WGMs with extremely narrow linewidths (high quality factor) \cite{vogt2018ultra}. The WGMR is mounted on a computer-controlled translation stage to adjust the position of the resonator relative to the waveguide; this allows for precise manipulation of the evanescent coupling efficiency of the WGMs, which is essential to achieve near-critical coupling, as discussed below. Since the WGMs are sensitive to environmental temperature and humidity fluctuations, the entire setup is in a thermally insulated enclosure with a relative humidity of less than 10\,$\%$ \cite{foreman2015whispering}. The temperature of the resonator is actively controlled using Peltier elements, reaching typical temperature stability better than $\pm$5\,mK over 48 hours. The thermal isolation housing sits on an active vibration isolation table to reduce coupling to seismic noise in the lab.

\begin{figure}[htb]
\centering\includegraphics[width=0.48\textwidth]{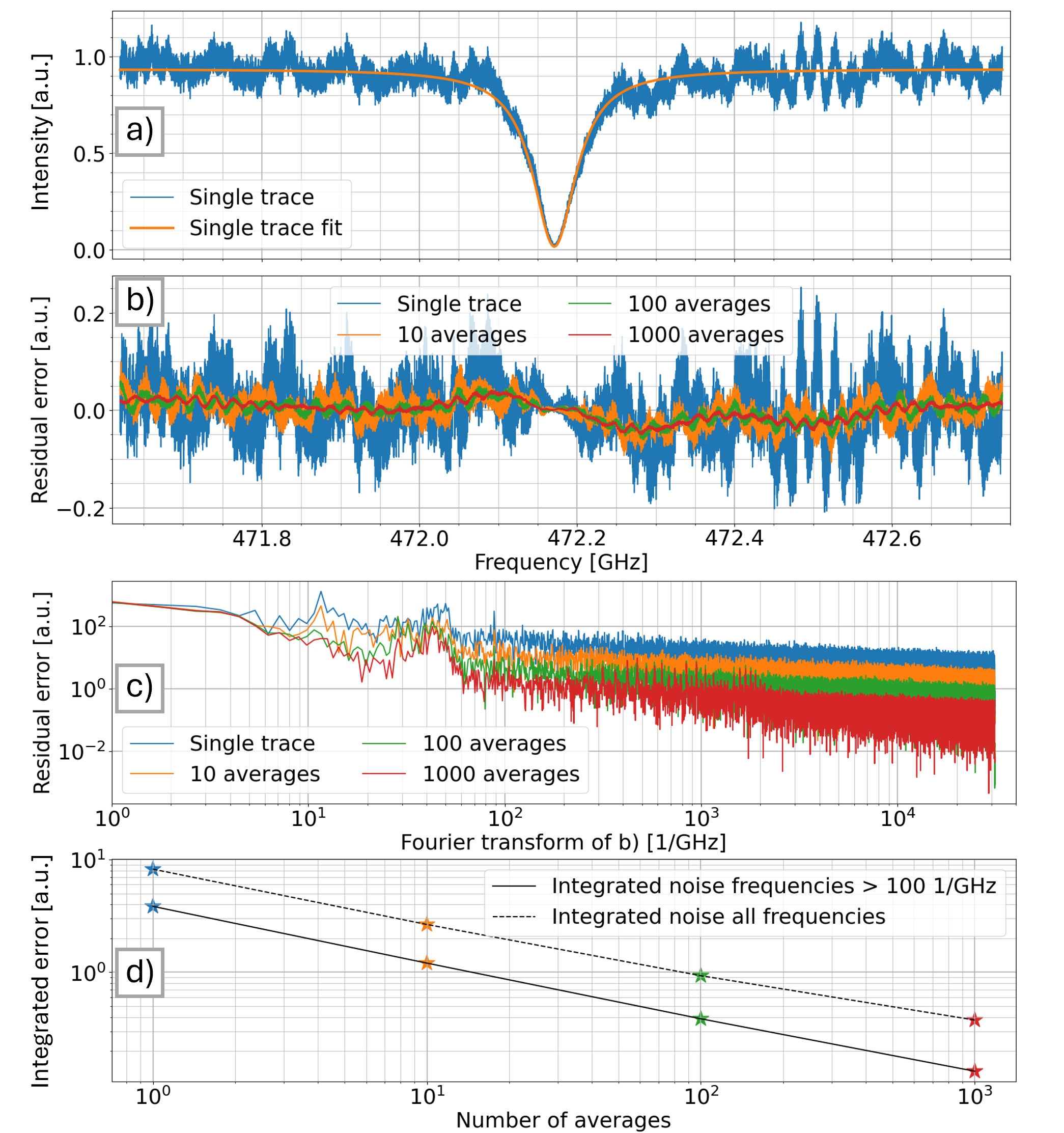}
\caption{(a) Single trace normalised waveguide transmission coupled to the WGMR (blue line) from 471.6\,GHz to 472.7\,GHz showing a single WGM, with the corresponding fit (orange line). (b) The residual error between the measurement and fitted analytical model for a single shot, 10, 100 and 1000 averaged traces. (c) Fourier analysis of the residua of fits shown in (b) to a WGM for different averaging. (d) Corresponding integrated residual noise (coloured stars) with close to $\sqrt n$ scaling. The data points are connected to guide the eye.}
\label{fig:2}
\end{figure}

The photocurrent at the PCA receiver is recorded using a high-resolution oscilloscope. The signal is frequency modulated due to the imbalanced THz interferometer and analysed with a Hilbert transform data analysis procedure. The Hilbert analysis allows retrieval of the amplitude and phase information from the photocurrent with a frequency resolution only limited by the linewidth of the THz laser source \cite{vogt2019coherent}. To characterise the THz WGMs, we normalise the transmission of the waveguide coupled to the WGMR (sample scan) to the transmission without a resonator (reference scan). This allows for full characterisation of the WGMs while minimising the impact of standing waves in the spectroscopy setup.  

Please note the zero offset of the difference frequency is calibrated to the mode order by aligning the tunable laser with the fixed frequency laser. Residual frequency errors are well below the stability of the experimental WGMR setup. The absolute stability and calibration of the difference frequency scan achieve measurement time-limited signal-to-noise (SNR) when averaging multiple scans \cite{RobinsonTait:2019,Puppe:xx}. A typical normalised waveguide transmission when coupled to the WGMR is shown in Fig. \ref{fig:2}. Increasing the number of averages clearly improves the achievable SNR.

\section{Results}

\begin{figure}[b]
\centering\includegraphics[width=0.45\textwidth]{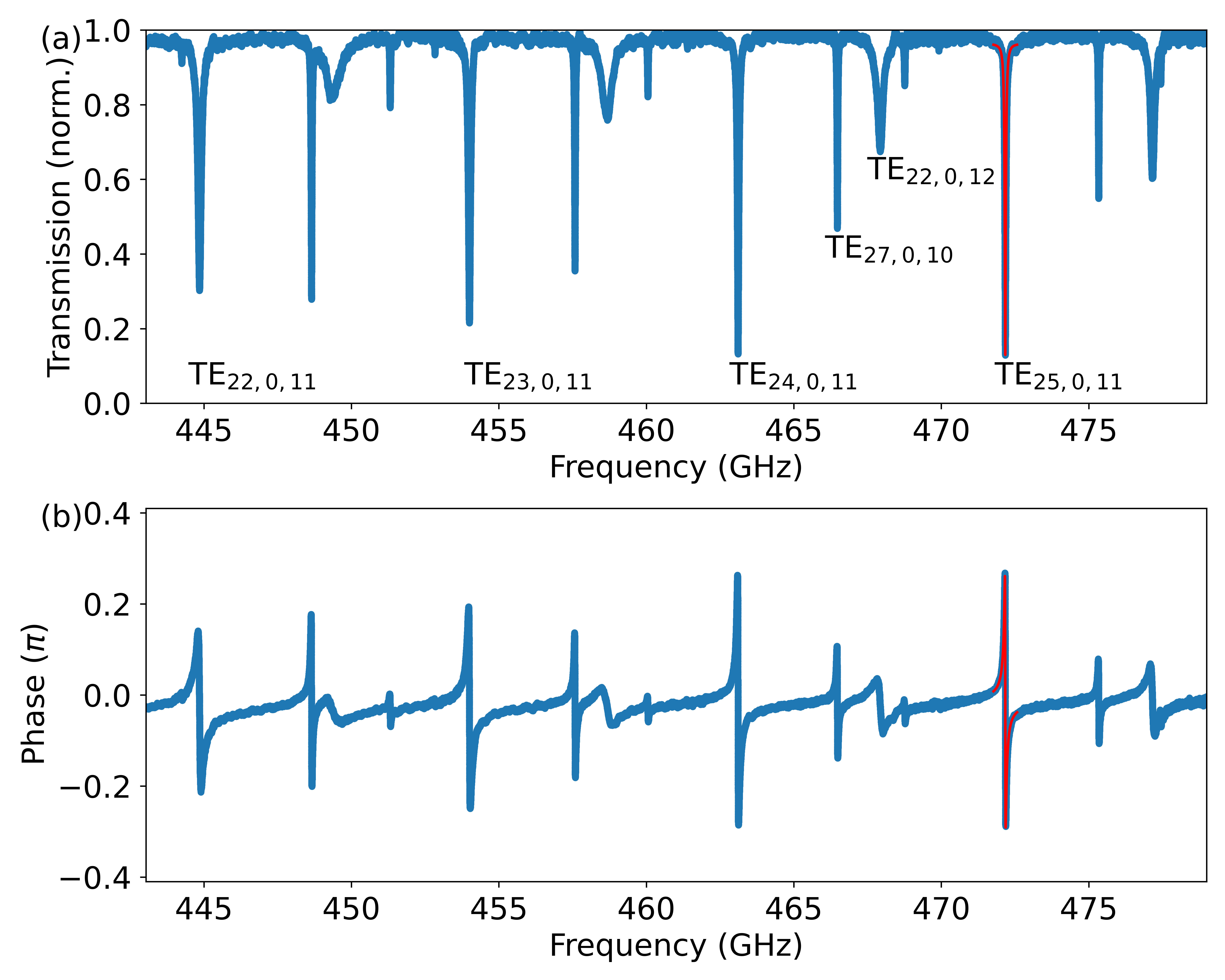}
\caption{a) Normalised transmission of the air-silica step-index waveguide coupled to 4\,mm diameter spherical silicon microresonator in the frequency range from 443 GHz to 479 GHz, normalised
to the waveguide transmission without the microresonator. (b) The corresponding phase profile to (a). The highlighted mode in red is shown in Fig. \ref{fig:4}.}
\label{fig:3}
\end{figure}

Typical measurements of the normalised transmission and phase of the WGMR in the frequency range from 443\,GHz to 479\,GHz over several free spectral ranges (FSRs) of the WGMR are shown in Fig. \ref{fig:3} (a) and (b), respectively. Four main mode families, each with their particular FSR, can be identified by comparison to Comsol Multiphysics\textsuperscript{\tiny\textregistered} finite element simulations. The excited WGM are transverse electric (TE) modes and are labelled according to the index in azimuthal (m), polar direction (p) and radial (q). For example, mode ${\mathrm{TE}}_{24,0,11}$ has 24 wavelengths around the circumference of the WGMR, a single maximum in polar direction and 11 maxima in radial direction \cite{breunig2013whispering}. 

The scan in Fig. \ref{fig:3} encompasses 2.5 million points over 40\,GHz leading to a 20\,kHz resolution. The exceptional absolute frequency stability of the FDS facilitates highly reproducible scans and allows the averaging of multiple scans. In Fig. \ref{fig:3}, 150 scans are averaged to improve the SNR (as discussed above). The acquisition time per scan is about 5\,s corresponding to a scanning speed of 8\,GHz per second.

\begin{figure}[bht]
\centering\includegraphics[width=0.45\textwidth]{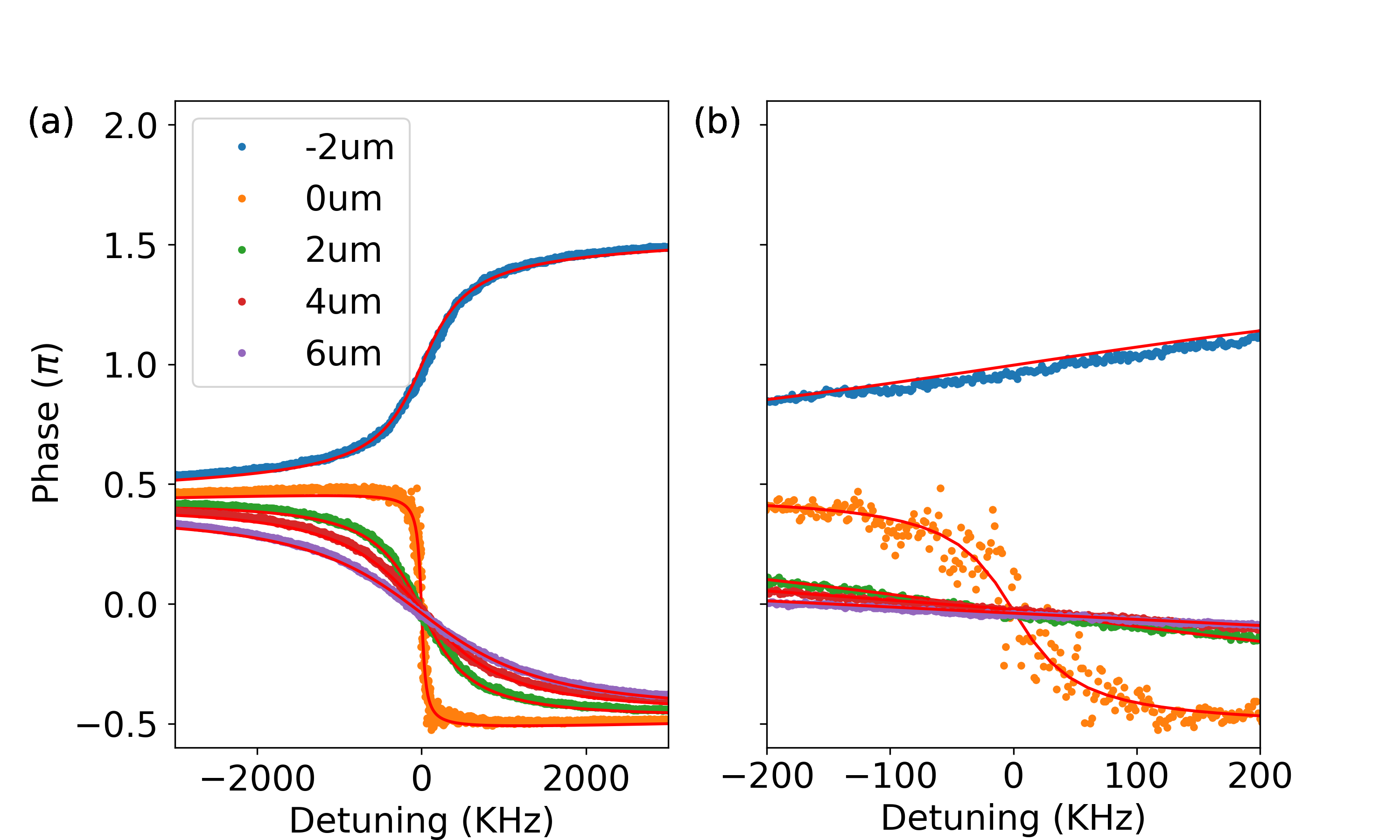}
\caption{(a) Phase profile for five different distances with $\pm$ 2000\,kHz detuning around the resonance frequency at 472.2\,GHz (${\mathrm{TE}}_{25,0,11}$). (b) Same measurements as in (a) but with a detuning of 200\,kHz to highlight the steep phase profile close to the resonance frequency.}
\label{fig:4}
\end{figure}

The spacing between the WGMR and the waveguide is chosen such that the modes are under-coupled, as can be seen from their corresponding phase profile \cite{vogt2018ultra}. At critical coupling, the phase profile features a $\pi$ phase jump at the resonance frequency \cite{vogt2017high}. Resolving this step is an ideal gauge for the effective frequency resolution of the FDS and the stability of the system. To this end, we optimise the distance between WGMR and waveguide to achieve close to critical coupling for the mode at 472.2\,GHz (${\mathrm{TE}}_{25,0,11}$; highlighted in red in Fig. \ref{fig:3}). The corresponding phase profiles for five different distances are shown in Fig. \ref{fig:4} (a), and zoomed in with $\pm$ 200\,kHz detuning around the resonance frequency in Fig. \ref{fig:4} (b). The WGM is overcoupled at a relative distance of \mbox{-2\,$\upmu$m} and eventually becomes undercoupled as the distance between the waveguide and the WGMR increases. At a relative spacing of 0\,$\upmu$m, the WGM is very close to critical coupling, with a relative width of the step-function of less than 200\,kHz, demonstrating the exceptional frequency resolution and stability of the novel FDS. Please note that the expected frequency resolution of the THz system itself is estimated to be several kHz, which can be further improved by locking the frequency comb to an optical reference. This being said, minute fluctuations in the WGMR's environment can lead to a broadening of the averaged phase transition and impede precise control of the coupling position e.g. due to thermal expansion and the thermo-optical coefficient of silicon \cite{vogt2018thermal}.




\begin{figure*}
\centering
\subfloat{\includegraphics[width=.8\linewidth]{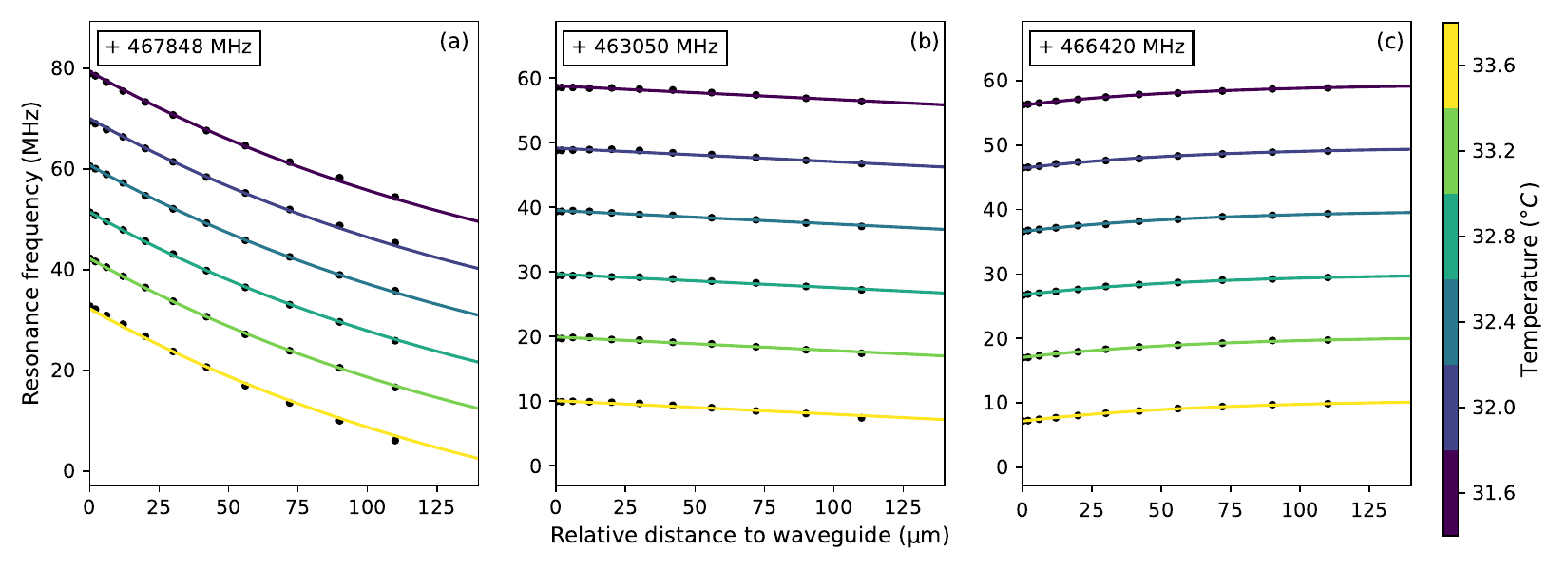}}
\\
\vspace*{-0.4cm}
\subfloat{\includegraphics[width=.8\linewidth]{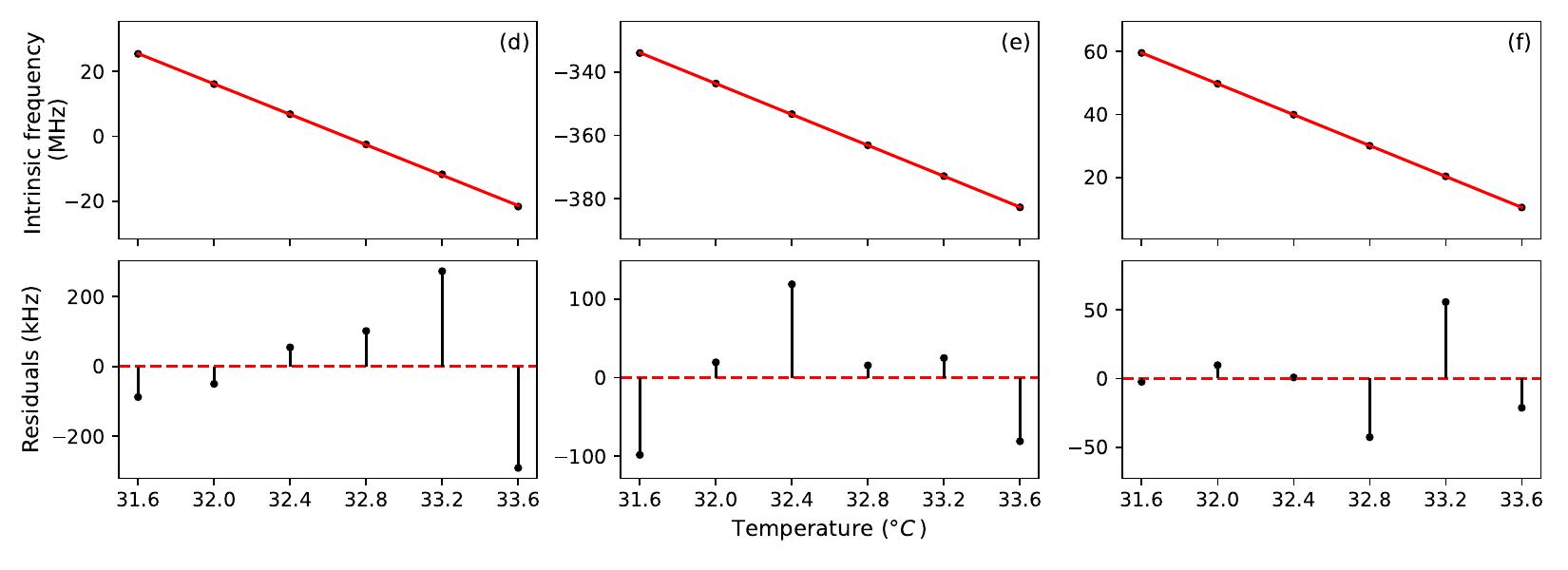}}
\vspace*{-0.4cm}
\caption{(a) to (c) show the measured resonance frequency shifts of modes ${\mathrm{TE}}_{22,0,12}$, ${\mathrm{TE}}_{24,0,11}$ and ${\mathrm{TE}}_{27,0,10}$ respectively (black dots), by changing the separation between the WGMR and the waveguide by about 110\,$\upmu$m. Each measurement is repeated at 31.6\,$^\circ C$, 32.0\,$^\circ C$, 32.4\,$^\circ C$, 32.8\,$^\circ C$, 33.2\,$^\circ C$ and 33.6\,$^\circ C$, and follows exponential behaviour (colour-coded lines). For clarity, the resonance frequencies are plotted with an offset as indicated in the top left corner of each subplot. (d) to (f) show the extracted intrinsic resonance frequencies (black dots) as a function of WGMR temperature with the corresponding linear fits (red lines). The residuals are shown below each subplot.}
\label{fig:5}
\end{figure*}


In addition to the exceptional frequency resolution, the FDS also has outstanding frequency stability. To further experimentally verify the performance of the FDS, we measure the frequency shifts of the WGMs as the relative spacing between the WGMR and the waveguide is altered. The waveguide's presence in the evanescent field of the WGMs is sufficient to shift the resonance frequencies by a few MHz. This is due to the change in the refractive index in the WGMs' environment. Figure \ref{fig:5} (a) to (c) shows exemplary the resonance frequency shifts of modes ${\mathrm{TE}}_{22,0,12}$, ${\mathrm{TE}}_{24,0,11}$ and ${\mathrm{TE}}_{27,0,10}$ for a relative waveguide WGMR spacing of about 0\,$\upmu$m to 140\,$\upmu$m, with 0\,$\upmu$m arbitrarily defined as the closest measured position. Note each measurement is repeated for six different temperatures (vertically offset curves), which will be discussed in detail below. The resonance frequencies at each position are extracted by simultaneously fitting an analytical model to the normalised transmission and phase profile of the WGMs \cite{vogt2018thermal}. The relative frequency shifts (black dots) follow an exponential trend, which can be seen from the very good agreement with the fit (colour-coded lines). Interestingly, mode families ${\mathrm{TE}}_{m,0,12}$ and ${\mathrm{TE}}_{m,0,11}$ experience a blue shift, while mode family ${\mathrm{TE}}_{m,0,10}$ a red shift. Notably, there is a transition from a strong blue shift (${\mathrm{TE}}_{22,0,12}$) to a red shift (${\mathrm{TE}}_{27,0,10}$) with mode ${\mathrm{TE}}_{24,0,11}$ experiencing a slight blue shift. Comparison with 3D finite element simulations reveals that mode families ${\mathrm{TE}}_{m,0,12}$ and ${\mathrm{TE}}_{m,0,11}$ have a lower effective refractive index ${n}_{\mathrm{eff}}$, and mode family ${\mathrm{TE}}_{m,0,10}$ a higher ${n}_{\mathrm{eff}}$ than the coupling waveguide. Also, albeit larger, mode family ${\mathrm{TE}}_{m,0,11}$ has an ${n}_{\mathrm{eff}}$ very similar to the waveguide. Accordingly, a mode with lower ${n}_{\mathrm{eff}}$ than the coupling waveguide is blue-shifted, while a mode with an ${n}_{\mathrm{eff}}$ larger than the coupling waveguide is red-shifted. To the best of our knowledge, this is the first time this behaviour of WGMs has been observed. 


As discussed above, we also analyse the temperature tuning of the WGMs, where each frequency shift measurement due to the presence of the waveguide is repeated for six temperatures from 31.6\,$^\circ C$ to 33.6\,$^\circ C$ in 0.4\,$^\circ C$ steps. At each temperature, the intrinsic resonance frequency of the WGMs is extracted from the exponential fits (resonance frequency at an infinitely large waveguide resonator distance). The corresponding intrinsic resonance frequencies for WGMs ${\mathrm{TE}}_{22,0,12}$, ${\mathrm{TE}}_{24,0,11}$ and ${\mathrm{TE}}_{27,0,10}$ are plotted in Fig. \ref{fig:5} (d) to (f), respectively. As expected, the intrinsic resonance frequencies follow a linear trend and are red-shifted with an increase in WGMR temperature \cite{vogt2018thermal}. The linear fits (red solid lines) show an excellent agreement with the data. The corresponding residuals are plotted below each graph. They are in the order of a few tens of kHz, clearly highlighting the frequency-domain spectrometer's exceptional frequency stability and reproducibility. Those results are at least two orders of magnitude better than typical existing commercial systems based on optical difference frequency generation and pave the way for highly sensitive sensors \cite{vogt2017high}.

\section{Conclusion}
The presented results clearly highlight the potential of the novel THz frequency-domain spectrometer based on a comb-locked frequency synthesizer. The spectrometer provides unprecedented opportunities for THz spectroscopy applications due to its capacity for kHz-level resolution and frequency stability while maintaining a bandwidth of several THz with scanning speeds of $>1$~THz/s. In particular, in combination with the ultra-high-Q THz WGMRs, a highly selective and sensitive spectrometer for THz sensing applications can be easily envisioned \cite{galstyan2021detection,yu2021whispering}. Moreover, referencing the novel spectrometer to a GPS-disciplined oscillator allows for reproducible results across different laboratories. This provides a significant advantage for high-precision THz spectroscopy. Our results also revealed a blue and red frequency shift of the WGMs due to the presence of the dielectric coupling waveguide. This previously unobserved behaviour provides novel insights into the intriguing physics of THz WGMs.


\section*{Data Availability Statement}
The data that support the findings of this study are available from the corresponding author upon reasonable request.

\nocite{*}
\bibliography{sample}

\end{document}